\documentclass[twocolumn]{revtex4}
\usepackage{graphicx}
\begin{document}
\baselineskip=12pt
\def\be{\begin{equation}}
\def\ee{\end{equation}}
\def\bea{\begin{eqnarray}}
\def\eea{\end{eqnarray}}
\def\E{{\rm e}}
\def\bearst{\begin{eqnarray*}}
\def\eearst{\end{eqnarray*}}
\def\peleven{\parbox{11cm}}
\def\peffec{\peight{\bearst\eearst}\hfill\peleven}
\def\pspace{\peight{\bearst\eearst}\hfill}
\def\ptwelve{\parbox{12cm}}
\def\peight{\parbox{8mm}}

\title
{Uncertainty in the Fluctuations of the Price of Stocks}

\author{G. R. Jafari$^{1}$, M. Sadegh Movahed $^{1}$, P. Noroozzadeh $^2$, A.
Bahraminasab $^3$,
\\Muhammad Sahimi $^4$, F. Ghasemi $^5$ and M.
Reza Rahimi Tabar$^{6,7}$}

\address{$^1$Department of Physics, Shahid Beheshti University,
Tehran 19839, Iran\\
$^4$Department of Physics, University of Antwerp,
Groenenborgerlaan 171, B-2020 Antwerpen, Belgium\\
$^3$Department of Physics, Sharif University of Technology,
Tehran 11365-9161, Iran\\
$^3$Physics Department, University of Lancaster, Lancaster, LA1
4YB United Kingdom\\
$^4$Mork Family Department of Chemical Engineering \& Materials
Science, USC, Los Angeles, California 90089-1211, USA\\
$^5$ The Max Planck Institute for the Physics of Complex Systems,
Nöthnitzer Strasse 38, 01187 Dresden, Germany\\
 $^6$Department of Physics, Sharif University of Technology,
Tehran 11365-9161, Iran\\
$^7$CNRS UMR 6202, Observatoire de la C$\hat o$te d'Azur, BP 4229,
06304 Nice Cedex 4, France}

\vskip 1cm


\begin{abstract}
We report on a study of the Tehran Price Index (TEPIX) from 2001 to
2006 as an emerging market that has been affected by several
political crises during the recent years, and analyze the
non-Gaussian probability density function (PDF) of the log returns
of the stocks' prices. We show that while the average of the index
did not fall very much over the time period of the study, its
day-to-day fluctuations strongly increased due to the crises. Using
an approach based on multiplicative processes with a detrending
procedure, we study the scale-dependence of the non-Gaussian PDFs,
and show that the temporal dependence of their tails indicates a
gradual and systematic increase in the probability of the appearance
of large increments in the returns on approaching distinct critical
time scales over which the TEPIX has exhibited maximum uncertainty.

Pacs{89.65.Gh, 89.75.-k }

\end{abstract}
\maketitle

\section{Introduction}
 In recent years, financial markets have been
a focus of physicists' attempts for applying the existing knowledge
from statistical mechanics to economic problems [1-3]. The markets,
though largely varying in the details of their trading rules and the
traded goods, may be characterized by some generic features of the
time series that describe the fluctuations in the prices of various
stocks and commodities. An important and challenging problem is to
understand and evaluate risk in the markets, which must be done
through the analysis of such time series. The aim of the analysis is
to characterize the statistical properties of the time series, with
the hope that a better understanding of its underlying stochastic
dynamics would provide useful information that can be used for
creating new models, that are able to reproduce experimental facts
(i.e., the actual recorded prices and their fluctuations).

A considerable amount of data and numerous studies indicate the
possibility that the financial time series may exhibit
self-similarity (and/or self-affinity) at short time scales which,
however, apparently breakdown at much longer times. Such features
are usually modeled in terms of various statistical distributions
with truncated tails. Recent studies indicated, however, that an
approach based on the Brownian motion [5,6], or other more
elaborated descriptions, such as those based on the L\'evy and
truncated L\'evy distributions [1], may not be suitable for properly
describing the statistical features of the fluctuations in the
stocks' price. Such models have been constructed based on the
premise that the financial time series may be viewed as {\it
additive} processes that are built up over time. There is now
increasing evidence that an approach based on {\it multiplicative}
processes might be a more  fruitful way of pursuing an accurate
analysis of the financial time series. This approach lends itself in
a natural way to multifractality [7] (see below). Such an idea was,
in fact, suggested some years ago when the intermittency phenomenon
in the returns fluctuations was observed at different scales, which
motivated some efforts for establishing a link between analysis of
the financial time series and other areas of physics, such as
turbulence [8-11]. We remind the reader that, if $p_i$ represents
the value of a stochastic variable at (time) $i$, the returns $r_i$
are defined by, $r_i=\ln(p_{i+1}/p_i)$. Nowadays, however, we know
that there are important differences between the two phenomena, such
as, for example, the differences between their spectra of
frequencies.

Based on the recent efforts for characterization of the various
stages of the development of markets [12-14], it is clear that
Tehran stock exchange represents an emerging market. It has
witnessed considerable activities over the past several years, but
it is still far from an efficient and developed market. Over a
two-year period, it lost more than 30\% of its value (from 13750
units in September 2004 to 9150 units in August 2006) and, on
average, the price of the stocks' units has decreased from \$0.92 to
\$0.49 (which, percentage-wise, represents an even steeper decline
than that of the units by which the market has lost value), even
without considering the rate of inflation. In addition, over the
past six months alone (up to the time of writing this paper), the
volume and values of the traded stocks have decreased by more than
60\%. Compared with the $S\& P$ 500, Tehran stock exchange is still
not a completely developed market [13], with its index exhibiting
stronger non-stationary features.

In this paper, we provide comprehensive evidence of the existence of
distinct critical time scales over which the Tehran Price Index
(TEPIX) has exhibited maximum uncertainty. Moreover, at several
critical times over the past few years, Tehran stock exchange has
been affected rather strongly by several political crises. These
features provide a good opportunity to test the method of analysis
suggested by Kiyono {\it et al.} [15] for an emerging market. More
specifically, by analyzing the temporal evolution of the index
dynamics, we demonstrate the strongly the non-Gaussian behavior of
the logarithmic returns of the TEPIX and scale-dependent behavior
(data collapse) of their probability density function (PDF). The
critical time scales are found to be in the vicinity of large index
movements, consistent with the high probability of multiscale events
at the critical points. From the observed non-Gaussian behavior of
the index, we numerically estimate the unexpectedly high probability
of a large price change near the critical times. Such estimates are
of importance to risk analysis, as they represent a central issue
for the understanding of the statistics of price changes.

The rest of this paper is organized as follows. In the next section
we present the data that we consider and describe how we analyze
them. The conclusions are summarized in Section 3.

\begin{figure}[t]
\includegraphics[width=9cm]{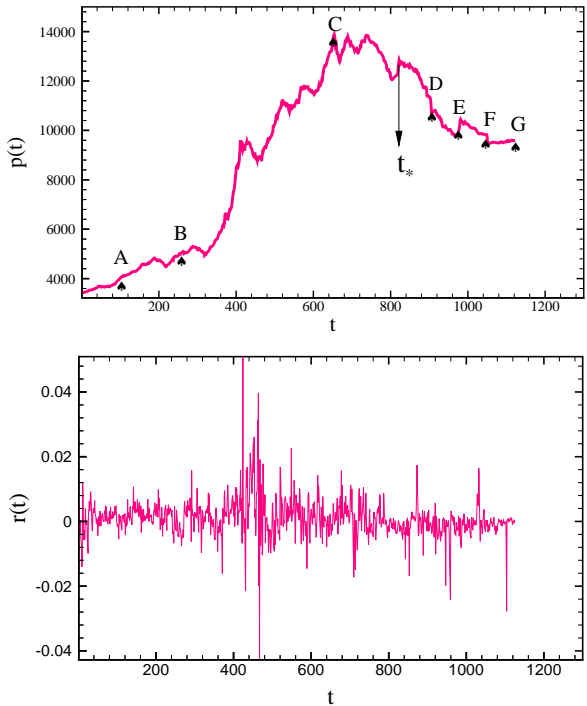} \narrowtext \caption{Top: history
(2001-2006) of daily deflated closure of TEPIX. Bottom: one day log
returns of the Tepix index.}\label{fig1}
\end{figure}
\begin{figure}[t]
\includegraphics[width=9cm]{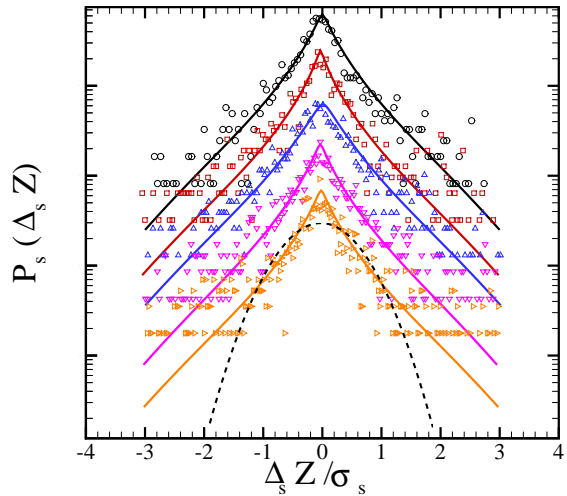} \narrowtext \caption{Continuous
formation of increment probability distribution function's across
scales for, from top to bottom, $s=4,8,12,16$ and 20 days. The solid
lines are the approximated PDF based on Casting's equation, the
right-hand side of Eq (4).}\label{cast}
\end{figure}

\begin{figure}[t]
\includegraphics[width=8cm]{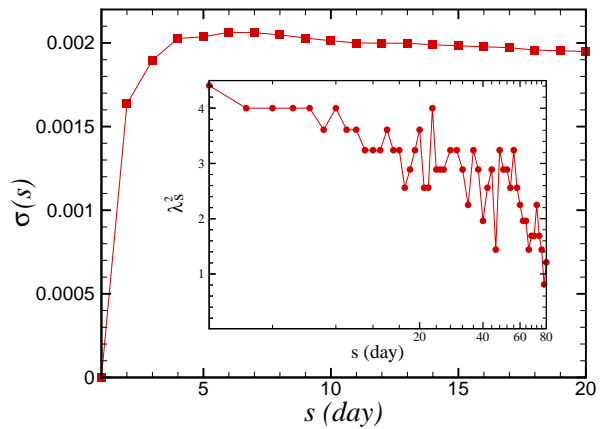} \narrowtext \caption{The
scale-dependence of the fitting parameter of Castaing's equation
$\lambda^2$ vs $\log s$. The inset shows the results for
$\sigma(s)=\sqrt{\langle (r(t)-\overline{r})^{2}\rangle}$. The
results indicate that after $s=4$ days, there is crossover in the
behavior of $\sigma$ vs $s$.} \label{cast1}
\end{figure}

\begin{figure}[t]
\includegraphics[width=8cm]{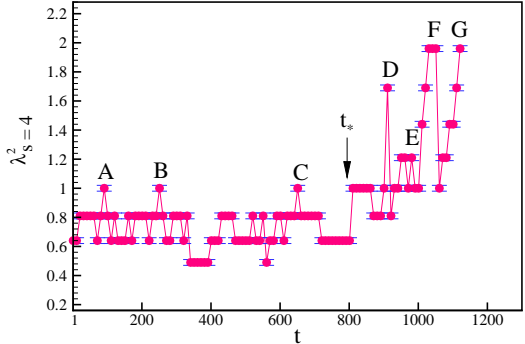} \narrowtext \caption{The local
temporal variation of $\lambda^2_{\rm 4\; days}$ over a one-year
period shows a gradual, systematic increase on approaching the
critical time scales A-G.}\label{cast1}
\end{figure}

\section{Analysis of the data}

Figure 1 shows the TEPIX over a period of over $4\;\frac{1}{2}$
years, from December 20, 2001 to August 10, 2006. The data had been
recorded on each trading day. We show in the lower panel of Fig. 1
the one-day log returns, i.e., $r_s(t)=\ln[p(t+s)/p(t)]$, where
$s=1$ day. We then analyze the PDF of the detrended log returns over
different time scales. To remove the trends present in $\{x(t)\}$,
where $x(t)=\ln p(t)$, we fit $x(t)$ in each subinterval $[1+s(k-1),
s(k+1)]$ of length $2s$ (where $k$ is the index of the subinterval
to a linear function of $t$ that represents the exponential trend of
the original index in the corresponding time window. After the
detrending procedure, we define detrended log returns on a scale $s$
as $\Delta_s p(t)= x^*(t+s)-x^*(t)$, where $1+s(k-1)\leq t\leq sk$,
and $x^*(t)$ is the deviation from the fitting function [7].

The scale-invariance properties of a fractal function $\Delta_s
p(t)$ are generally characterized by exponents $\xi_q$ that govern
the power-law scaling of the absolute moments of its fluctuations,
i.e., $m(q,l)=K_ql^{\xi_q}$, where, for example, one may choose
$m(q,l)=\sum_t|\Delta_s p(t+l)-\Delta_s p(t) |^q $. As is
well-known, if the exponents $\xi_q$ are linear in $q$, a single
scaling exponent $H$ suffices for characterizing the fractal
properties with, $\xi_q=qH$, in which case $\Delta_s p(t)$ is said
to be monofractal. If, on the other hand, the function $\xi_q$ is
not linear in $q$, the process $\Delta_s p(t)$ is said to be
multifractal. Some well-known monofractal stochastic processes are
self-similar processes with the following property, \be
\Delta_{\lambda s} p(t)=\lambda^H\Delta_s p(t),\hskip 1cm \forall
s,\; \lambda > 0\;. \ee Widely-used examples of such processes are
the fractional Brownian motion and the L\'evy walk. One reason for
their success is, as it is generally the case in experimental time
series, that they do not involve any particular scale ratio [i.e.,
there is no constraint on $s$ or $\lambda_s$ in Eq. (1)].

In the same spirit, one can try to build multifractal processes that
do not involve any particular scale ratio. A common approach,
originally proposed in the field of fully-developed turbulence
[8,15-18], has been to describe such processes in terms of
stochastic equations, in the scale domain, describing the cascading
process that determines how the fluctuations evolve when one passes
from the coarse to fine scales. One can state that the fluctuations
at scales $s$ and $\lambda_s$ are related (for fixed $t$) through
the cascading rule, \be \Delta_{\lambda s}p(t)=W_\lambda\Delta_s
p(t),\hskip 1cm \forall s,\;\lambda>0\;, \ee where $\ln(W_\lambda)$
is a random variable. Let us note that Eq. (2) can be viewed as a
generalization of Eq. (1) with $H$ being stochastic. Since Eq. (2)
can be iterated, it implicitly forces the random variable
$W_\lambda$ to have a log infinitely-divisible law [19]. It has been
demonstrated by Castaing {\it et al.} [19] that a non-Gaussian PDF
with ``fat'' tails can be modeled by random multiplicative
processes.

Thus, let us assume that the increments in the time series are
represented by the following multiplicative form [7]: \be \Delta_s
p(t)=\zeta_s(t)\exp[\omega_s(t)]\;, \ee where $\zeta_s$ and
$\omega_s$, assumed to be independent variables, are both Gaussian
random variables with zero mean and variances $\sigma_s^2$ and
$\lambda_s^2$, respectively. The PDF of $\Delta_s p(t)$ has fat
tails, depending on the variance of $\omega_s$, and is expressed by
[19]: \be\label{castaing} P_s(\Delta_s p)=\int
F_s\left(\frac{\Delta_s p}{\sigma_s}\right) \frac{1}{\sigma_s}
G_s(\ln \sigma_s)d\ln\sigma_s\;, \ee where we have assumed that
$F_s$ and $G_s$ are both Gaussian with zero mean and variance
$\sigma_s$ and $\lambda_s$, i.e., \be
G_s(\ln\sigma_s)=\frac{1}{{\surd 2\pi}\lambda_s}\exp\left(-
\frac{\ln^2 \sigma_s}{2\lambda^2_s}\right)\;. \ee Thus, we may
investigate the time scale-dependence of $\lambda^2_s$. In this
case, the equation for $P_s(\Delta_s p)$ is referred to as
Castaing's equation, whose solution converges to a Gaussian when
$\lambda\to 0$.

The fit of the PDF of TEPIX increments to Castaing's equation is
indeed almost perfect, especially within $\pm 3$ standard
deviations, even for a single record. This is demonstrated in Fig.
2. Although Eq. (4) is equivalent to that for a log-normal cascade
model - originally introduced to study fully-developed turbulence
[19] - it approximately describes the non-Gaussian PDFs observed not
only for turbulence, but also in a wide variety of other phenomena,
ranging from rate of exchange of foreign currencies [8], to
heartbeat interval fluctuations [15,20]. Also shown in Fig. 2 is the
fit of the data for $s=20$ days to a Gaussian distribution, which
clearly fails to represent the data.

For a quantitative comparison, we fit the data (over the
$4\;\frac{1}{2}$ years interval) to the above function [Eq. (4)], as
illustrated in Figs. 2 and 3, and estimate the variance
$\lambda_s^2$ of $G_s$. As shown in Fig. 3, the standardized
(variance = 1) PDF of the detrended log returns indicates the
existence of a scaling law in the behavior of $\lambda_s^2$ as a
function of $s$, rather than logarithmic decay which is
characteristic of classical cascade processes [17-19,21]. Figure 3
indicates that, after $s=4$ days, there is a crossover in the
behavior of $\lambda_s$ as a function of $s$. For comparison, we
have also calculated the variance $\langle r(t+\tau)r(t)\rangle$
(which represents the width of the joint probability distribution).
The results are shown in the inset of Fig. 3. Similar to
$\lambda_s$, there is a crossover in the behavior of width.

In the following, we identify a temporal region of complete
departure from the cascade scenario to an instance of the
critical-like behavior. We evaluate (in sliding time intervals
$[t-\Delta t/2, t+\Delta t/2]$) the temporal dependence of
$\lambda_s^2$. The local temporal variation of $\lambda_{s=4\;{\rm
days}}^2$ over a one-year period shows a gradual, systematic
increase on approaching the critical time scales A-G identified in
Fig. 4. It it beneficial to risk analysis to quantify the
non-Gaussian nature of (detrended) price fluctuations on a
relatively short time scale ($\sim$ 4 days), and not just the
volatility at larger time scales [1], which is what is normally
analyzed. The important point is that large values of $\lambda_s^2$
indicate a high probability of a large price change; this
probability follows a sharp increase with growing $\lambda_s^2$.

The critical points are denoted by $A$ to $G$ in Fig. 4. To plot the
Fig. 4 we chose a moving window with length $\Delta t = 150$ days.
It may be interesting to note that these time scales are related to
the political developments in Iran. There was an increasing trend in
the price index over the time scale A, caused by privatization of
Iran's industries. B represents the time period from February 21,
2003, when the inspectors of the International Atomic Energy Agency
(IAEA) and its Director-General, Dr. Mohammed ElBaradei, travelled
to Iran, to June 16, 2003, when Dr. ElBaradei reported to the IAEA's
Board of Governors on what the IAEA had found in Iran. C represents
the restart by Iran of production of centrifuges' parts, used in
uranium enrichment (UE), on July 31, 2004. D is the time period that
included the European Union's warning to Iran that it would cut off
the negotiations on May 11, 2005; Iran's subsequent declaration on
May 19, 2005 that its UE program is irreversible, and the election
of Iran's new president on June 26, 2005. E is the time period over
which a new director for Tehran stock market was appointed, and the
economic policies of Iran's new president were declared. Finally, F
is the time at which the IAEA reported to the United Nations
Security Council Iran's nuclear dossier on February 27, 2006. The
time $t_*$ represents the time at which Iran's rejection of the IAEA
demand for stopping work on the construction of a heavy-water
nuclear reactor in Arak was announced on February 13, 2005. It can
be seen clearly in Fig. 4 that, after that time the TEPIX entered a
critical period that has continued up to now. Moreover, as Fig. 4
indicates, similar to most major stock markets around the world, the
Tehran stock market has responded almost immediately to the
political events on the dates indicated. As shown in Fig. 4, the
trends in the TEPIX are essentially stable up to time scale $C$, but
beyond C the average uncertainly increases.

To check the changing of the nature of fractal distribution of the
returns, we plot (in a semi-logarithmic graph) the PDFs of the
one-day returns before and after the critical time $t_*$. The
results are presented in Fig. 5. Relative to a Gaussian
distribution, they exhibit sharp peaks, but not long tails. In Table
1, we compare the means, standard deviations, skewnesses, and
kurtosises of the returns time series before and after the time
$t_*$, as given in Fig. 1. As Table 1 indicates, the mean value of
the returns is negative after $t_*$, but positive before $t_*$.
Moreover, the variance after $t_*$ is smaller that its value before
$t_*$, implying that, on average, the investors have lost their
investments after $t_*$ but, with smaller risk, had gained before
$t_*$.

\begin{figure}[t]
\includegraphics[width=10cm]{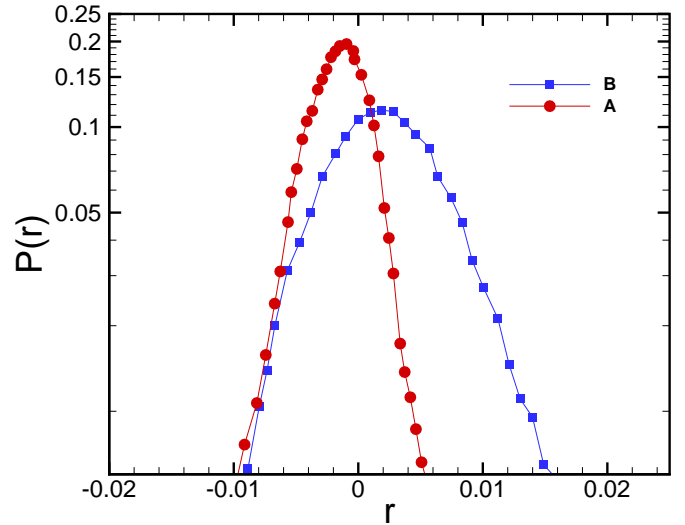} \narrowtext \caption{(Color
online)Probability distribution function of the TEPIX returns before
time B and after time A $t_*$.} \label{fig2}
\end{figure}

\begin{table}[htb]
\begin{center}\label{Tb1}
\caption{Comparing the general parameters of the one-day return
series.}
\begin{tabular}{|c|c|c|c|c|}
& Mean & Standard Deviations & Skewness & Kurtosis \\ \hline $t<t_*$
& $ 0.00172$ & $0.006$ & $0.84$ & $16.65$ \\ \hline
$t>t_*$ & $-0.00104$ & $0.004$ & $-0.66$ & $12.46$\\
\end{tabular}
\end{center}
\end{table}
\section{Summary}
Tehran stock exchange provides a good opportunity to test the
recently developed analyzing method suggested by kiyono \emph{et al}
in an emerging market. We characterized the non-Gaussian nature of
the detrended log returns of the Tehran price index from 2001 to
2006 using a model based on multiplicative processes, and found the
empirical evidence that the temporal dependence of fat tails in the
PDF of the detrended log returns shows a gradual, systematic
increase in the probability of the appearance of large increments,
on approaching distinct critical time scales. The results suggest
the importance of the non-Gaussian behavior at a time scale (4 days)
for risk analysis. If the same characteristics are observed in other
stock indices, our approach may be applicable to quantitative risk
evaluation.

\section{Acknowledgment}

We thank Didier Sornette for useful comments and discussions.


\begin{thebibliography} {50}

\bibitem{1} J.-P. Bouchaud and M. Potters, {\it Theory of Financial Risks, from
Statistical Physics to Risk Management} (Cambridge University Press,
London, 2000).

\bibitem{2} D. Sornette, Phys. Rep. 378 (2003) 198.

\bibitem{3} D. Sornette, {\it Why Stock Markets Crash?} (Princeton University
Press, Princeton, 2003).

\bibitem{4} G. Brumfiel, Nature 435 (2005) 132.

\bibitem{5} R. Cont and J.-P. Bouchaud, Macroecon. Dyn. 4 (2000) 170.

\bibitem{6} T. Lux and M. Marchesi, Nature 397 (1999) 498.

\bibitem{7} K. Kiyono, Z. R. Struzik, and Y. Yamamoto, Phys. Rev. Lett. 96
(2006) 068701.

\bibitem{8} S. Ghashghaie, W. Breymann, J. Peinke, P. Talkner, and Y. Dodge, Nature
381 (1996) 767.

\bibitem{9} P. Jefferies, M. L. Hart, P. M. Hui, and N. F. Johnson, preprint,
cond-mat/9910072.

\bibitem{10} D. Challet {\em et al.}, Quant. Fin. 1 (2001) 168.

\bibitem{11} J. Davoudi and M. R. Rahimi Tabar, Phys. Rev. Lett. 82 (1999) 1680.

\bibitem{12} G. R. Jafari, M. S. Movahed, S. M. Fazeli, M. R. Rahimi Tabar, J. Stat.
Mech. (2006) P06008.

\bibitem{13} P. Noroozzadeh and G. R. Jafari, Physica A 356 (2005) 609.

\bibitem{14} G. R. Jafari, A. Behraminasab and P. Noroozzadeh, Accepted in Int. J. Mod. Phys. C. (2007), physics/0503027.

\bibitem{15} K. Kiyono, Z. R. Struzik, N. Aoyagi, S. Sakata, J. Hayano, and Y.
Yamamoto, Phys. Rev. Lett. 93 (2004) 178103.

\bibitem{16} H. E. Stanley and V. Plerou, Quant. Fin. 1 (2001) 563.

\bibitem{17} J. F. Muzy, J. Delour, and E. Bacry, Eur. Phys. J. B 17 (2000) 537.

\bibitem{18} A. Arneodo, E. Bacry, S. Manneville, and J. F. Muzy, Phys. Rev. Lett.
80 (1998) 708.

\bibitem{19} B. Castaing, Y. Gagne, and E. J. Hopfinger, Physica D 46 (1990) 177.

\bibitem{20} K. Kiyono, Z. R. Struzik, N. Aoyagi, F. Togo, and Y. Yamamoto, Phys. Rev.
Lett. 95 (2005) 058101.

\bibitem{21} B. Chabaud, A. Naert, J. Peinke, F. Chilla, B. Castaing, and B. Hebral,
Phys. Rev. Lett. 73 (1994) 3227.


\end{thebibliography}
\end{document}